\documentclass[lettersize,journal]{IEEEtran}

\usepackage{tabularx}

\ifCLASSINFOpdf
\else
\fi
\usepackage[numbers]{natbib}
\usepackage{epsfig}
\usepackage{epstopdf}
\usepackage[T1]{fontenc}
\usepackage{amssymb}
\usepackage{amsmath}
\usepackage{amsfonts}
\usepackage{verbatim}
\usepackage{graphicx}
\usepackage{hyperref}
\usepackage[usenames,dvipsnames]{xcolor}
\usepackage{lipsum}
\usepackage[ruled,vlined,linesnumbered]{algorithm2e}

\SetKwComment{Comment}{\tiny{//}}{}

\usepackage{array}
\newcolumntype{P}[1]{>{\centering\arraybackslash}p{#1}}
\newcolumntype{M}[1]{>{\centering\arraybackslash}m{#1}}
\usepackage{multirow}
\usepackage[caption=false,font=normalsize,labelfont=sf,textfont=sf]{subfig}
\usepackage{multicol}
\usepackage{pgfplots}
\usepackage{lipsum}
\usepackage{colortbl}
\usepackage{colortbl}
\usepackage{tabularx}
\usepackage{lipsum}  
\usepackage{pifont}
\usepackage{makecell}
\usepackage{float}
\newcommand\subsubsubsection{\@startsection{paragraph}{4}{\z@}%
  {1.5ex \@plus 1ex \@minus .2ex}%
  {-1em}%
  {\normalfont\normalsize\bfseries}}
\makeatother

\usepackage{tikz,bm,color}

\usetikzlibrary{circuits.logic.US,circuits.logic.IEC,fit}
\newcommand\addvmargin[1]{
  \node[fit=(current bounding box),inner ysep=#1,inner xsep=0]{};
}
\begin{document}
%
\title{FastFlow in FPGA Stacks of Data Centers}

\author{\IEEEauthorblockN{Rourab Paul$^1$, Alberto Ottimo$^2$, Marco Danelutto$^3$}
\\
\IEEEauthorblockA{
Dept. of Computer Science, University of Pisa, Italy$^{1,2,3}$\\
rourab.paul@unipi.it$^1$, alberto.ottimo@unipi.it$^2$, marco.danelutto@unipi.it$^3$} 
}

\maketitle
\vspace{-10pt}

\begin{abstract}
FPGA programming is more complex as compared to Central Processing Units (CPUs) and  Graphics Processing Units (GPUs). 
The coding languages to define the abstraction of Register Transfer Level (RTL) in High Level Synthesis (HLS) for FPGA platforms have emerged due to the laborious complexity of Hardware Description Languages (HDL). 
The HDL and High Level Synthesis (HLS) became complex when FPGA is adopted in high-performance parallel programs in multicore platforms of data centers. Writing an efficient host-side parallel program to control the hardware kernels placed in stacks of FPGAs is challenging and strenuous. The unavailability of efficient high level parallel programming tools for multi core architectures makes multicore parallel programming very unpopular for the masses.
This work proposes an extension of $FastFlow$ where data flows in hardware kernels can be executed efficiently in FPGA stacks. 
Here host side codes are generated automatically from simple $csv$ files. The programmer needs to specify four simple parameters in these $csv$ file: FPGA IDs, source, destination nodes, hardware kernel names. The proposed tool flow uses $FastFlow$ libraries with $Vitis$ to develop efficient and scalable parallel programs for FPGA stacks in data centers. The evidence from the implementation shows that the integration of $FastFlow$ with  $Vitis$ reduces $\sim$ 96 \% coding effort (in terms of number of lines) as compared to existing $Vitis$ solutions.
\end{abstract}
\begin{IEEEkeywords}
Structured Parallel Programming, Data Center, FPGA and FastFlow.
\end{IEEEkeywords}
\vspace{-5pt}
\section{Introduction}
The modern heterogeneous computing system consists of CPUs, GPUs, and FPGAs. The CPU platforms are the most flexible architecture to support libraries of compute efficient FPGAs and compute-dense GPUs. FPGAs are faster and more energy-efficient than GPUs and CPUs for high processing algorithms. However, the massive array of small processing units inside FPGA fabric makes FPGA programming more complex compared to CPUs and GPUs. In recent years, many programming models \cite{uhd}, \cite{panda} and \cite{vivado:hls} have been developed where high-level language is used for FPGA programming rather than low-level RTL codes. 
$Xilinx-AMD$ is the most popular FPGA manufacturer company that designed $Vitis$ \cite{vitis} unified software platform, which includes hardware kernel design, host side application development, analysis, and implementation on FPGA. However, designing host side parallel code for multiple FPGAs in $Vitis$ platform is very complex. The $Vitis$ GitHub \cite{vitis:accl} repository shows that, a simple data flow in hardware kernels needs few hundred lines of host-side code. 
To reduce the coding complexity of parallel programming in multi core platform, an open-source parallel programming framework, $FastFlow$ \cite{fastflow}, was developed through collaboration between University of Pisa and Torino. 
\vspace{-5pt}
\subsection{Existing Tool Flows}
\label{sec:toolflow}
The existing tool flows can be categorized in three directions.
\subsubsection{Existing $FastFlow$ without FPGA \citep{fastflow}, \cite{tonci}}
To reduce coding complexity in parallel programming, $FastFlow$ provides application programmers with pre-defined parallel programming patterns written in a C++ library. These patterns can be used to design complex parallel applications. The $FastFlow$ is based on four fundamental principles: (i) hierarchical design to support local optimization and incremental design, (ii) stream parallelism of data, (iii) improvement of the abstraction level for parallel programmers, and (iv) efficiency of base mechanisms. Though $FastFlow$ was originally designed for shared memory and multicores, it has been extended to support parallel applications on COW/NOW architectures \cite{tonci}. The $FastFlow$ primarily handles stream parallel patterns such as pipeline, task farm, and data parallel patterns like map, reduce, and stencil. It also supports higher-level and more general patterns such as divide and conquer, data flow, and sequential wrapper patterns, which make it appropriate for reusing existing business code of parallel computation. $FastFlow$ uses a wrapper named $ff\_node\_t<Tin,Tout>$ class, which runs inside a thread and processes $Tin*$ data and producing outputs  $Tout*$. It is to be noted that all the components of $FastFlow$ are represented as $ff\_node\_t$, which executes sequentially and wraps multiple staged pipelines, stencil computations, data parallelism, etc.
\subsubsection{Vitis Flow}
\label{sec:vitisflow}
The $Xilinx-AMD$ FPGA required two primary files for FPGA configurations.
The $xclbin$ generation (FPGA configuration file) process of $Vitis$ tool flow involves two steps. The first step generates $xo$ files of the hardware kernels required for the system by taking two types of input files: 1) source files of kernels written in HDL, CPP, or SystemC and 2) the connectivity files to describe the memory connection of input and output ports of the hardware kernels. These $xo$ files link hardware kernels with the target FPGA technology. In the second step, the $xo$ files of link hardware kernels are used to generate $xclbin$ file of targeted FPGA technology.
\begin{figure}[h]
\centering
\vspace{-5pt}
\includegraphics[width=0.35\textwidth]{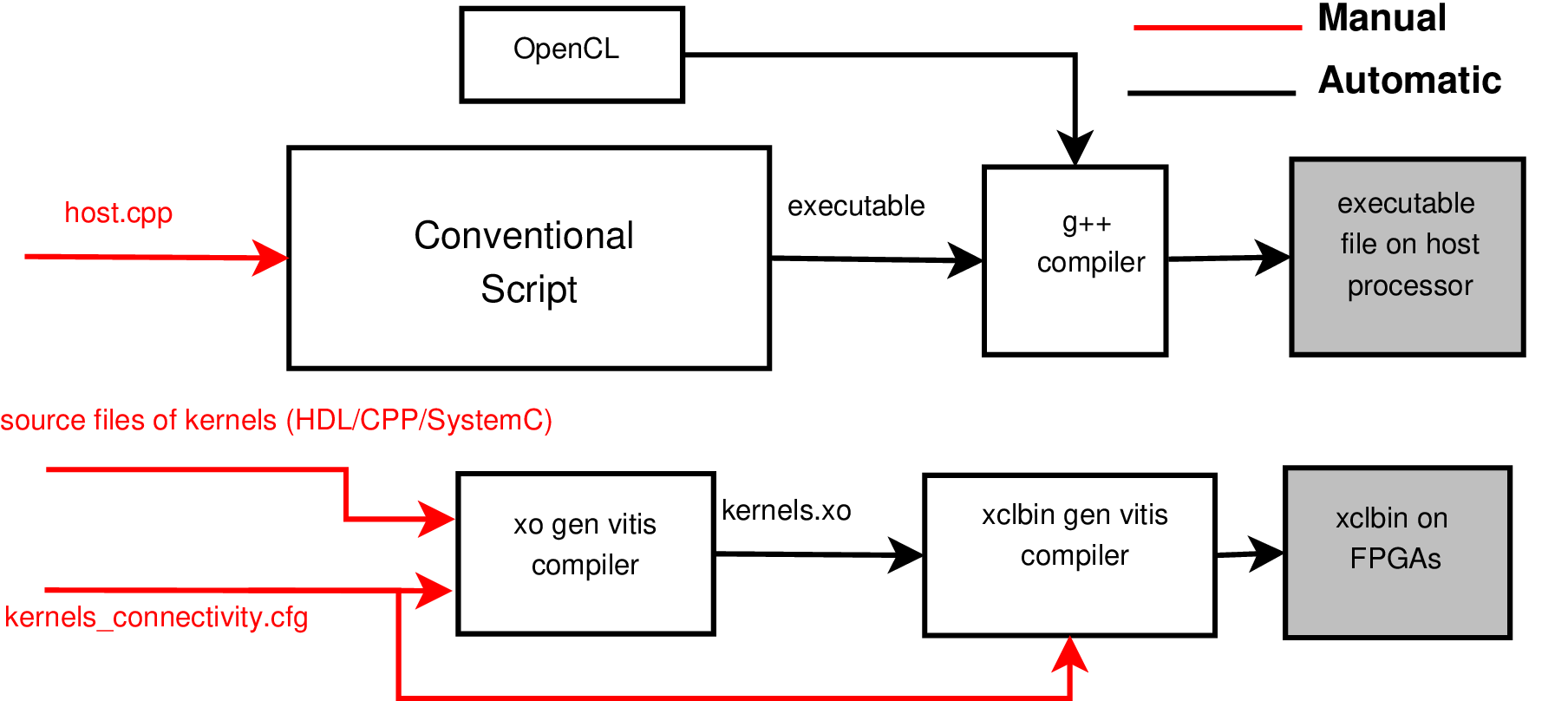}
\vspace{-5pt}
\caption{Vitis Tool Flow}
\vspace{-5pt}
\label{fig:vitis_flow}
\end{figure}
The $Vitis$ flow may require $OpenCL$ in some $host.cpp$ to load hardware kernels on FPGAs. The hardware programmer needs to write the code of the $host.cpp$ manually. $Vitis$ uses the $gcc$ compiler to generate an executable from $host.cpp$. The executable from $host.cpp$ runs on the host processor. The entire tool flow of $Vitis$ is shown in Fig. \ref{fig:vitis_flow} where complex $host.cpp$, memory connection files of hardware kernel ($connectivity.cfg$) needs to be written manually.

\subsubsection{Existing FastFlow + Vitis Flow \citep{ffpga}}
\begin{figure}[!htbp]
\vspace{-10pt}
\centering
\includegraphics[width=0.39\textwidth]{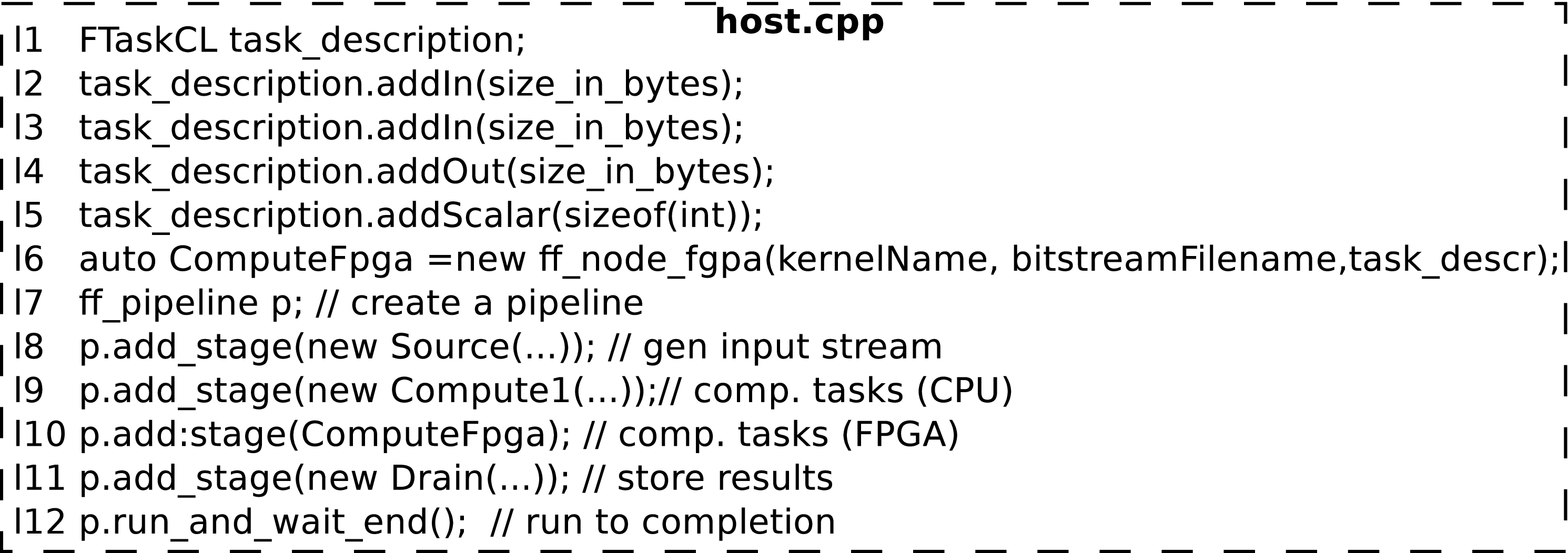}
\caption{host.cpp in Existing FastFlow + Vitis}
\vspace{-10pt}
\label{fig:ex_ff_vitis}
\end{figure}

The initial versions of $FastFlow$, as described in \cite{fastflow} and \cite{tonci}, execute task computation on the CPU using $ff\_node\_t$. However, the $Vitis$ integrated $FastFlow$, as mentioned in \cite{ffpga}, added a special $ff\_node\_fpga$ to execute task computation inside the FPGA. Execution of task in FPGA using $ff\_node\_fpga$ requires $xclbin$, hardware kernel names, inputs and outputs information of the kernels. The execution steps of  $ff\_node\_fpga$ involves with input and output transfer hardware kernels using $svc$ method and computation of hardware kernel. 
As depicted in Fig. \ref{fig:ex_ff_vitis}, line $6$ creates the $ff\_node\_fpga$ and line $10$ add the hardware kernel in specific pipeline stage. Line $8$, $9$ and $11$ add pipeline stages in CPU. Finally the line $12$ deploys it inside FPGA. This pipeline stages are replicable according to the structure  process flow.
\subsection{Our Contribution}
This manuscript develops a scalable multiple FPGA support in $FastFlow$ \cite{fastflow} environment where precompiled hardware kernels are already built by $Vitis$ Tool \cite{vitis}. The standard parallel patterns of tasks like pipelines and farms are provided by a simple $proc.csv$ file to the $FastFlow$ script. However, in the article \cite{ffpga}, instead of creating a simple $proc.csv$ file like ours, the designers need to develop the $host.cpp$ file from scratch. The proposed $FastFlow\_fpga\_stack\_script$ includes $FastFlow$ in $Vitis$ to generate an executable file to map the entire parallel patterns of tasks mentioned in $proc.csv$ in multiple heterogeneous or homogeneous FPGAs.  
The primary contributions of this work are stated below:

\begin{itemize}
\item This work proposes a hybrid tool flow with $FastFlow$, $OpenCL$ and $Vitis$ libraries to develop efficient and scalable parallel programs for a stack of FPGAs in data centers. This framework places hardware kernels in target FPGAs, ensuring appropriate data synchronization of their input-output ports.
\item Without incurring additional FPGA overhead, our scheme can generate complex parallel host-side code from a simple $csv$ file. 
\item The process flow of hardware kernels generated by our tool flow achieves the same performance as that generated from $Vitis$. Our framework is scalable for FPGA stacks and multiple hardware kernels, reducing coding effort by approximately $\sim$ 96\% (\# of lines) compared to existing $Vitis$ solutions.
\end{itemize}

\IEEEpeerreviewmaketitle

\section{Proposed FastFlow + Vitis for FPGA Stacks}
The proposed new $FastFlow$ + $Vitis$ Flow for FPGA Stacks has three primary differences compared to the existing $FastFlow$ + $Vitis$.
\subsection{Automation of Tool Flow}
To reduce the coding effort in FPGA stacks, the proposed $FastFlow$ integrated $Vitis$ hybrid tool flow automates the generation of connectivity files and $host.cpp$.
\subsubsection{xclbin \& Memory Connection Files Generation}
The $xclbin$ generation process in $FastFlow$ integrated $Vitis$ is similar to the conventional $Vitis$ Tool flow stated in Sec. \ref{sec:vitisflow} except the generation of connectivity files. Usually the memory connectivity files are written manually in  $Vitis$. Whereas $FastFlow$ integrated  $Vitis$ tool flow generates $xclbin$ and connectivity files of hardware kernels from $circuit.csv$. 
\subsubsection{host.cpp Generation}
\label{sec:fastflow}
\begin{figure*}[!htbp]
\centering
\vspace{-5pt}
\includegraphics[width=0.94\textwidth]{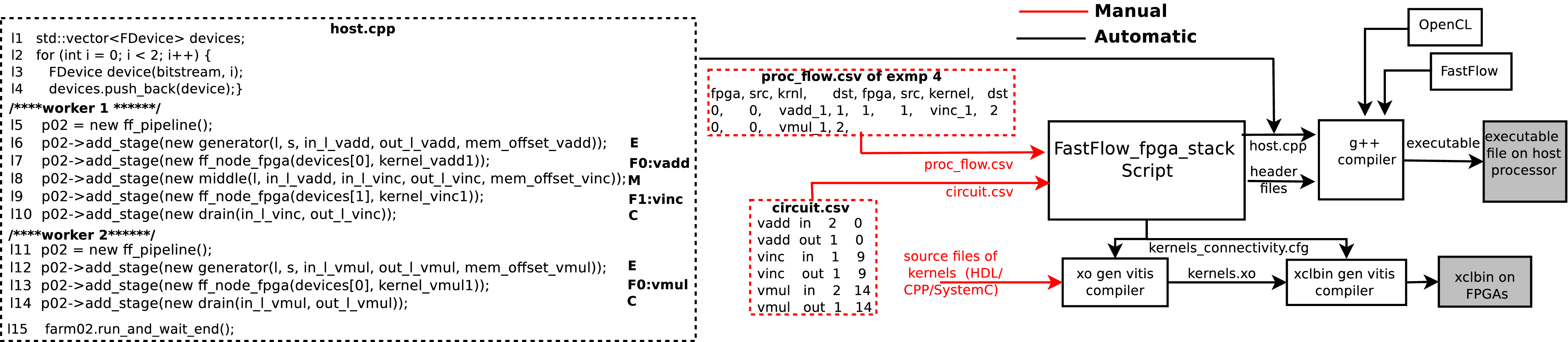}
\caption{Proposed Tool Flow of Fastflow Integrated Vitis}
\vspace{-10pt}
\label{fig:ff_vitis_flow}
\end{figure*}

The $FastFlow$ integrated $Vitis$ tool generates $host.cpp$ by taking $2$ input files: $proc.csv$ and $circuit.csv$. The $proc.csv$ is a list of hardware kernels with four fields: 1)$FPGA-ID$ to identify target FPGA from FPGA stacks where hardware kernel will run, 2) $Src$ node to connect inputs of hardware kernel, 3)$Dst$ node to connect outputs of hardware kernel and 4)$Kernel$ $Name$ to identify the require hardware kernel. If the $src$ nodes of multiple hardware kernels are the same, these kernels collect inputs from the same nodes. If the $dst$ nodes of one kernel match with the $src$ nodes of a different kernel, it indicates that the output of the first kernel is connected to the input of the next kernel. The $circuit.csv$ file describes the number of inputs, outputs, and memory slots used for the inputs and outputs of the hardware kernels. The entire tool flow of our framework is shown in Fig. \ref{fig:ff_vitis_flow}. The pseudo code of the proposed script is shown in Algo. \ref{alg:script}.

\begin{algorithm}[htb]
\small
\captionsetup{belowskip=-20pt} 
\caption{$FastFlow\_fpga\_stack\_script$}\label{alg:script}
\KwData{$proc.csv$, $circuit.csv$, $kernel$ $source$ $files$}
\KwResult{$host.cpp$, $connectivity.cfg$, $xclbin$}
WhitespaceFilter($proc.csv$, $circuit.csv$)\Comment*[m]{\tiny{white space remove}}
file\_rule\_check($proc.csv$, $circuit.csv$)\Comment*[m]{\tiny{check file format}}
\For{$i=1$ to $m$}{\tcp{\tiny{generate memory connection files for $m$ types kernels}}
$connectivity.cfg\_m$ = con\_gen($circuit.csv$)}
\tcp{\tiny{generate xclbin for $m$ types kernels}}
xclbin = bit\_gen($connectivity.cfg\_m$, kernel source files)\;
uq\_farms = find\_uq\_farms($proc.csv$)\Comment*[r]{\tiny{compute \# farm(s)}}
req\_fpga($proc.csv$)\Comment*[r]{\tiny{calculate required \# fpgas}}
\For{all unique\_farms }{
  \If{multiple workers}{\tcp{\tiny{for \# worker>1 with \# pipeline>1 \&\# pipeline=1}}
     \If{multiple pipes}{
         write\_multpipe\_farm($host.cpp$)\
      }
     \If{one pipe}{
     write\_onepipe\_farm($host.cpp$)\
     }    
  }
  \If{one worker}{\tcp{\tiny{for \# worker=1 with \# pipeline>1 \&\# pipeline=1}}
      \If{multiple pipes}{
         write\_multpipe\_1worker($host.cpp$)\;
      }
     \If{one pipe}{
     write\_onepipe\_1worker($host.cpp$)\;
     } 
     }
     \vspace{-8pt}
}

\end{algorithm}

\vspace{-5pt}
\subsection{Scalability for FPGA Stacks}
To make the $FastFlow$ integrated $Vitis$ tool scalable for FPGA stacks, two major changes are adopted in our $FastFlow$ library.
\subsubsection{FastFlow FPGA Node}
The construction of $ff\_node\_fpga$ is modified to pass the FPGA ID. 
As shown in $host.cpp$ of Fig. \ref{fig:ff_vitis_flow}, it creates a vector of $FDevice$ objects named $devices$. It then iterates $n$ times, where $n$ is the number of FPGA devices required to map the given process flow. The value of $n$ is determined by our proposed $FastFlow\_fpga\_stack\_script$. In Fig. \ref{fig:ff_vitis_flow}, for the given $example$ $4$, $n=2$. Lines $2$ to $3$ of $host.cpp$ in Fig. \ref{fig:ff_vitis_flow} create an $FDevice$ object named $device$ with parameters xclbin: $bitstream$ and $i$, and then add this $device$ to the $devices$.
\subsubsection{Pipeline Stages}
In our $FastFlow\_fpga\_stack\_script$, the architecture of the $add\_stage$ method \cite{ffpga} to add pipeline stages is modified. 
In our revised $FastFlow\_fpga\_stack\_script$, the $add\_stage$ method is updated to utilize the $devices$ vector and kernel name for adding $ff\_node\_fpga$. In contrast, the previous version of $FastFlow$+$Vitis$ used the $bitstream$ ($xclbin$), hardware kernel name and input/output parameters of the hardware kernel for the $add\_stage$ method.
In line $7$, $9$, and $13$ of $host.cpp$ (Fig. \ref{fig:ff_vitis_flow}), the kernel names : $vadd$, $vinc$, and $vmul$, along with the FPGA ID, are passed in $ff\_node\_fpga$.

\subsubsection{Nodes \& Parameters}
\begin{figure*}[!htbp]
\centering
\vspace{-10pt}
\includegraphics[angle=90, height=272pt, width=0.25\textwidth]{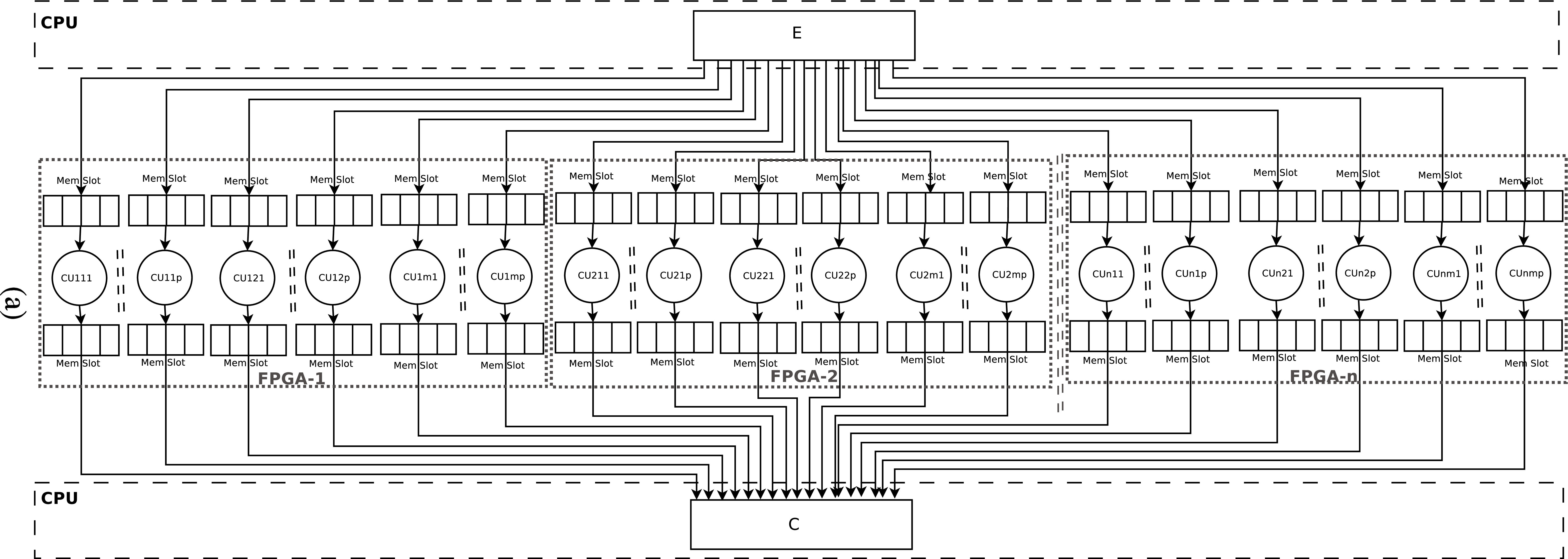}
\includegraphics[width=0.72\textwidth]{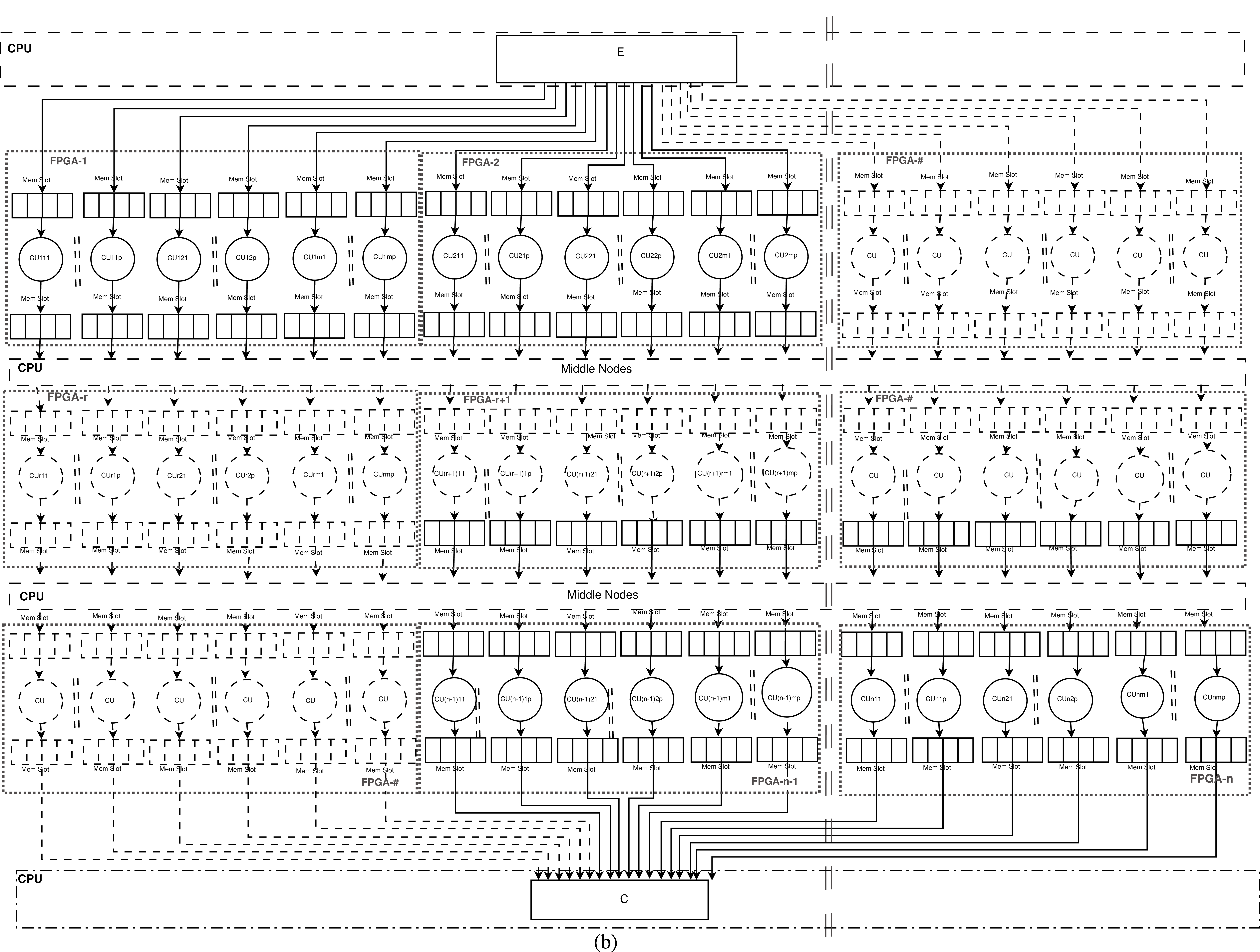}
\vspace{-5pt}
\caption{(a) Farms with 1 pipeline stage (b) Farms with multiple pipeline stages}
\vspace{-5pt}
\label{fig:pipe}
\end{figure*}

Similar to the previous $FastFlow$ version, the proposed toolflow also includes two primary components: 1) Hardware kernels and 2) Communication of hardware kernels using $svc$ method. The farms with 1 pipeline stage and farms with multiple pipeline stages  shown in Fig. \ref{fig:pipe}(a) and Fig. \ref{fig:pipe}(b) respectively are represented by three indexes $n$, $m$ and $p$ where $n$ is FPGA ID, $m$ describes hardware kernel types ($vadd$, $vinc$, $vmul$, etc.) and $p$ represents kernels indexes available in the FPGA. 
 All the hardware kernels have different numbers of input and output slots connected to the device's global memory through dedicated memory slots which may require to achieve the highest throughput. In our case, the memory slots are HBM, DRAM, and PLRAM. 
In our current version of the tool, four types of nodes are available as a pipeline stages: 1) Emitter ($E$), 2) Collector ($C$), 3) FPGA node ($F$) where hardware kernel (CU) is placed, and 4) Middle ($M$). Here, each node runs inside a thread.
The $E$ node connects data streams to hardware kernel inputs, while the $C$ node collects data from various kernel outputs. The $M$ nodes are required when hardware kernels outputs are connected with another hardware kernel's input. 
The $F$ node runs inside the FPGA, while the $E$, $C$, and $M$ nodes run within the CPU of the host processor. The $E$, $C$, $M$, and $F$ nodes are represented as $generator$, $drain$, $middle$, and $ff\_node\_fpga$ in Fig. \ref{fig:ff_vitis_flow}, respectively.

\section{Implementation and Result}
\begin{figure*}[!htbp]
\centering
\includegraphics[width=0.49\textwidth]{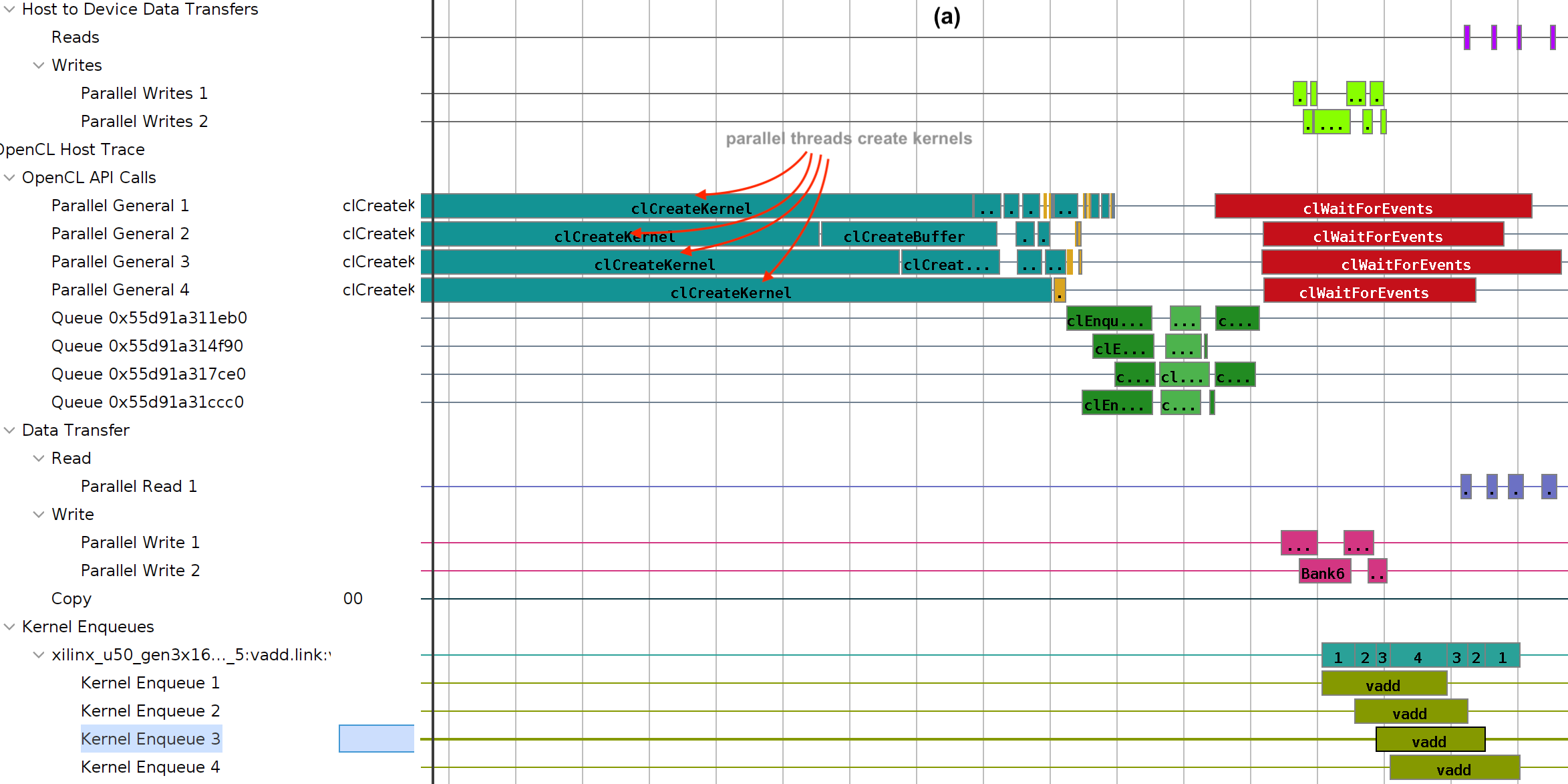}
\includegraphics[width=0.48\textwidth]{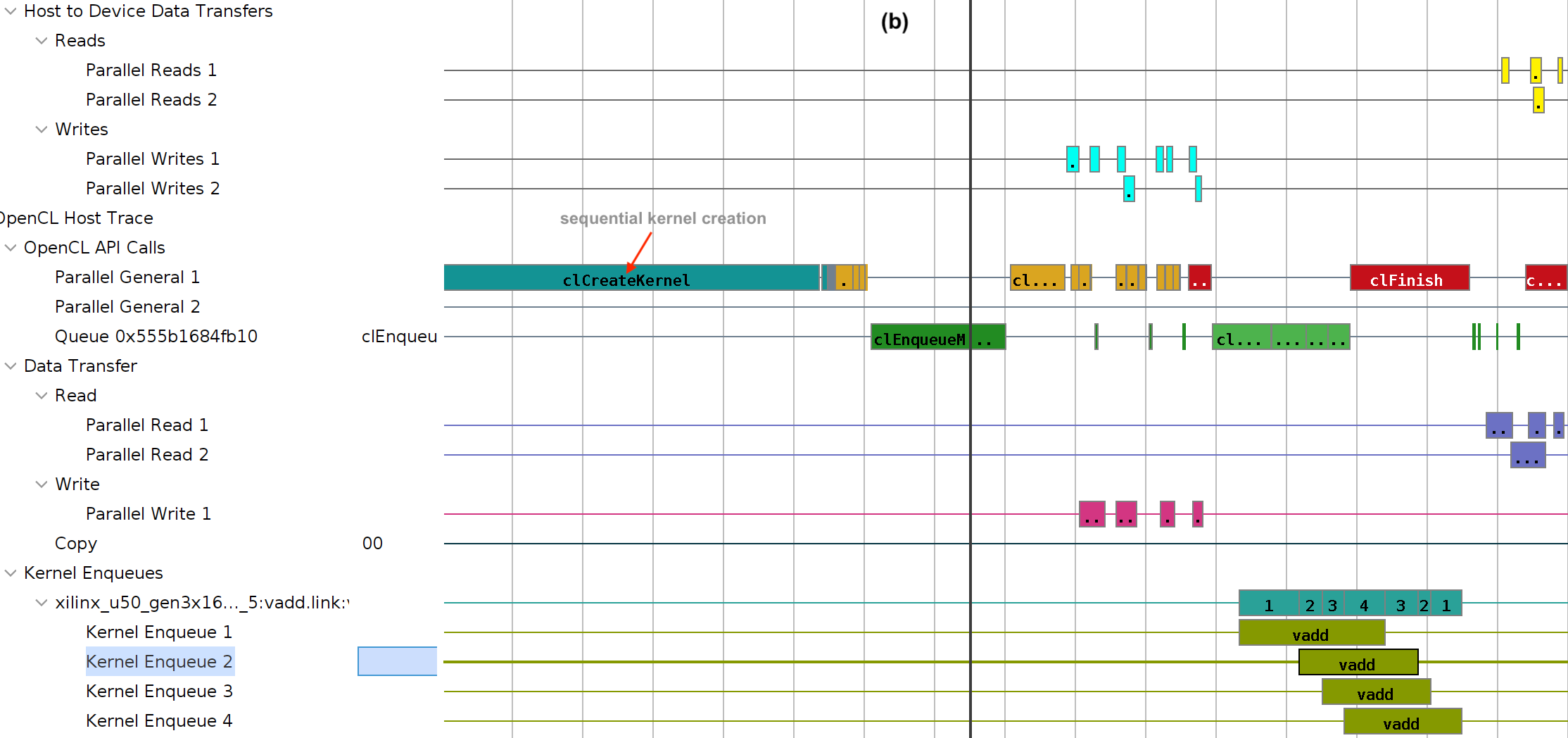}
\caption{Example 1 in Vitis Analyzer : (a) Proposed $FastFlow$+$Vitis$ vs (b) $Vitis$}
\vspace{-5pt}
\label{fig:exam1}
\end{figure*}

\begin{figure*}[!htbp]
\centering
\includegraphics[width=0.41\textwidth]{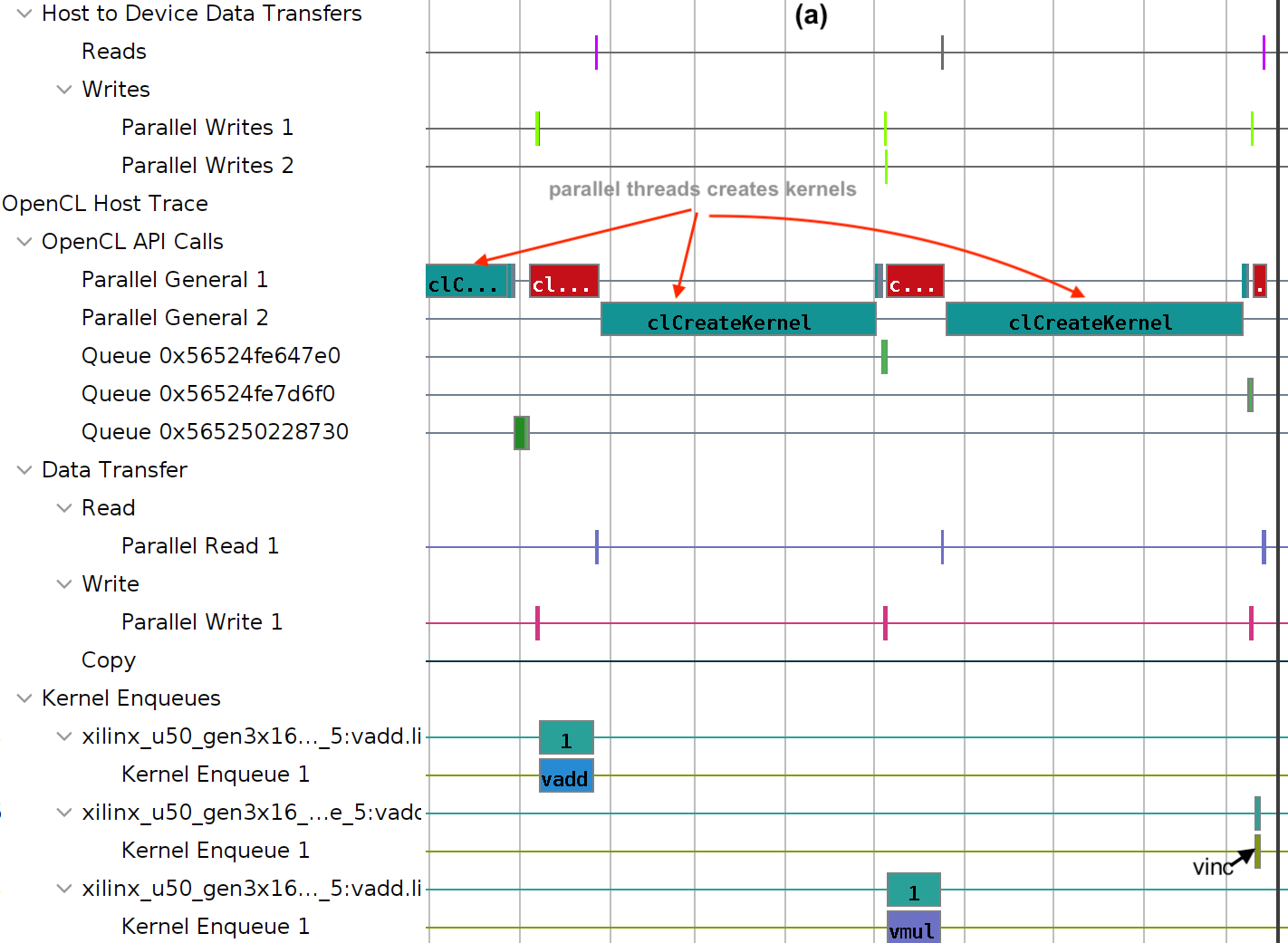}
\includegraphics[width=0.5\textwidth]{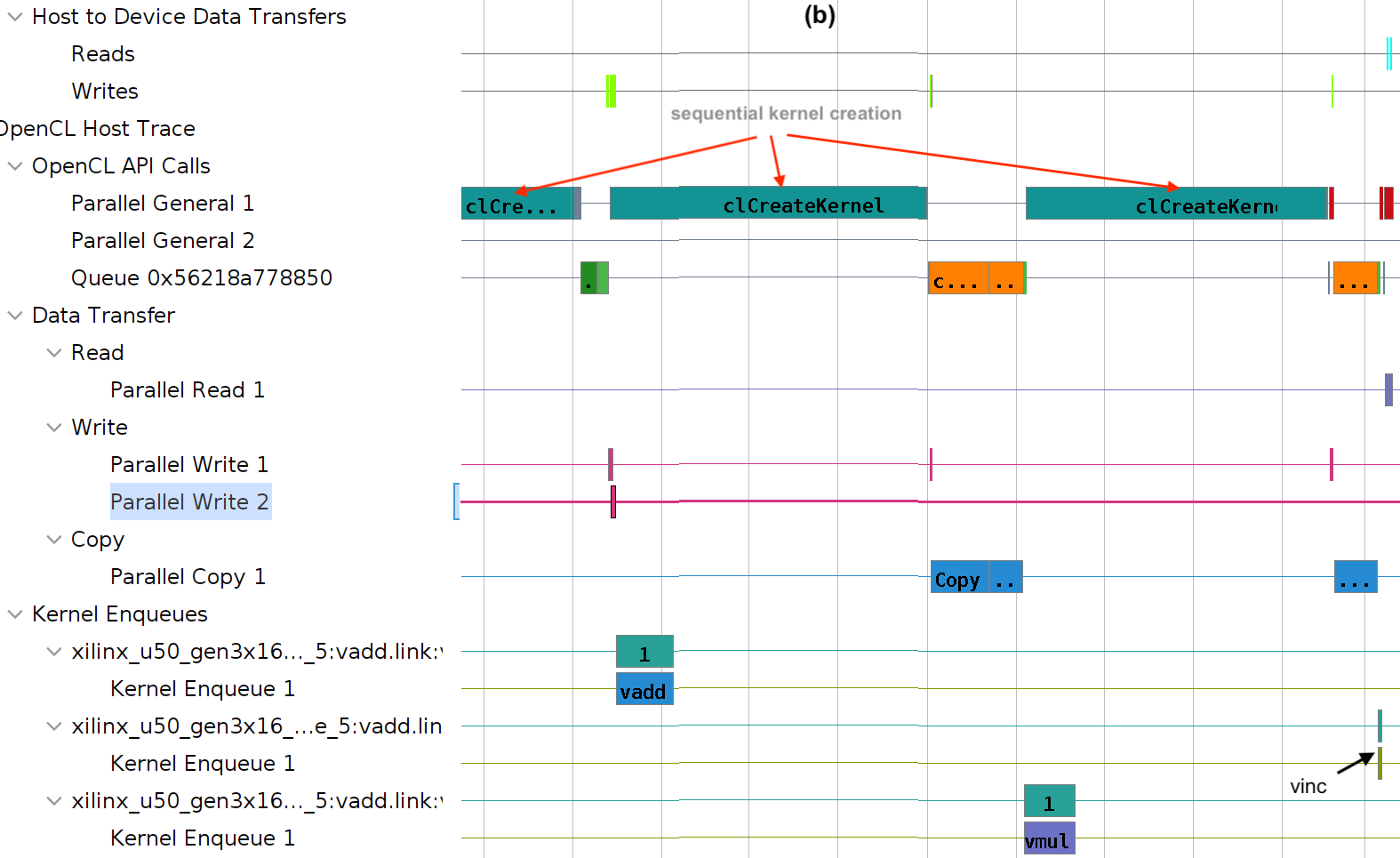}
\caption{Example 2 in Vitis Analyzer : (a) Proposed $FastFlow$+$Vitis$ vs (b) $Vitis$}
\vspace{-5pt}
\label{fig:exam2}
\end{figure*}

\begin{figure*}[!htbp]
\centering
\vspace{-5pt}
\includegraphics[angle=270, width=0.37\textwidth]{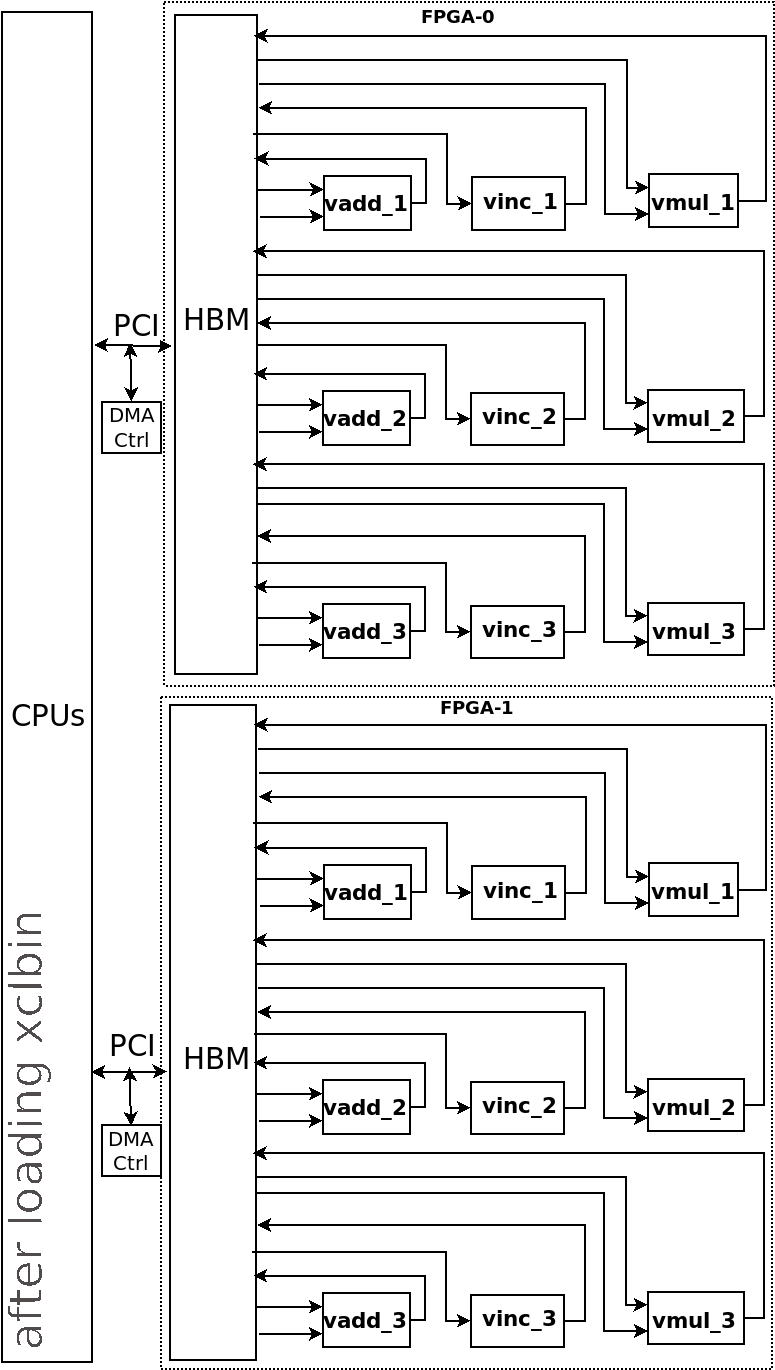}
\includegraphics[angle=270, width=0.39\textwidth]{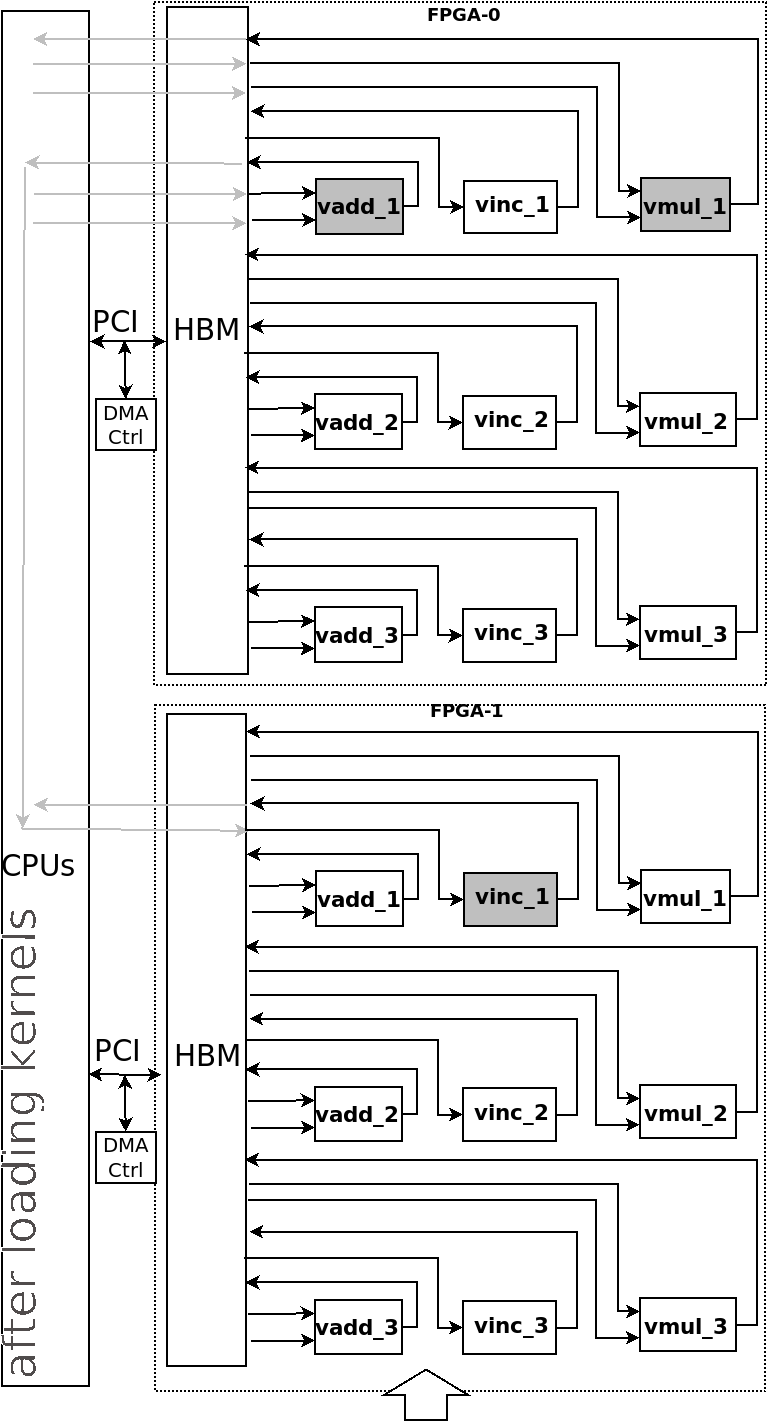}
\caption{Example 4 in 2 FPGAs With $FastFlow$+$Vitis$ : }
\label{fig:exm4}
\vspace{-5pt}
\end{figure*}

The proposed $FastFlow$ + $Vitis$ tool flow is implemented and tested on two $Xilinx-AMD$ $Alveo-U50$ FPGAs using $Vitis$ $2023$. This work reports five fundamental synthetic types of farms and pipes which may be used as basic blocks of a larger design. Example 1 shown in Table \ref{tab:comp} consists of four workers with vector addition ($vadd$) hardware kernels named $vadd\_1$, $vadd\_2$, $vadd\_3$, and $vadd\_4$, which can run parallelly in this farm. In $Vitis$ flow, the programmer needs to write $165$ lines of code in the $host.cpp$ and $8$ lines in connectivity file, whereas in our tool, the programmer only needs to write $6$ lines combined in the $proc.csv$ and $circuit.csv$. For example 1, our framework reduces the number of lines in $host.cpp$ by $65\%$ compared to the $Vitis$. Example 2, shown in Table \ref{tab:comp}, has one worker with three pipeline stages named $vadd\_1$, $vmul\_1$, and $vinc\_1$. In example 2 using the $Vitis$ flow, $host.cpp$ required $273$ lines, whereas in our $Vitis$ + $FastFlow$ combination, the programmer only needs to compose $1$ line in $proc.csv$ and $6$ lines in $circuit.csv$. For example 2, our framework reduces the number of lines in $host.cpp$ by $86\%$ compared to the $Vitis$ flow. 
Similarly, Table \ref{tab:comp} reports the same for other examples.
For other examples, 4$^{th}$ and 6$^{th}$ columns of Table \ref{tab:comp} report the number of lines required to write manually by the programmer for $Vitis$ and our $FastFlow$+$Vitis$ respectively. 
The power consumption, critical path and slice utilization are same across all five designs developed using $FastFlow$+$Vitis$ and $Vitis$. 
Fig \ref{fig:exm4} shows the connections of three hardware kernels of example 4: $vadd\_1$, $vinc\_1$  and $vmul\_1$ placed in $FPGA-0$ and $FPGA-1$. The gray color kernels get loaded after the execution of $host.cpp$ generated form our $FastFlow$+$Vitis$ script. 
The $host.cpp$ shown in Fig \ref{fig:ff_vitis_flow}, 5 lines from $2$ to $6$ represent 5 pipeline stages ($E$, $F0:vadd$, $M$, $F1:vinc$, $C$) of $worker-1$ and 3 lines from $8$ to $10$ represent 3 pipeline stages  ($E$, $F0:vmul$, $C$) of  $worker-2$  of example 4. In Fig. \ref{fig:exam1}, the $Vitis$ $analyzer$ compares the timing diagram of $example$ $1$ in both $Vitis$ and $Vitis$+$fastflow$. Similarly, in Fig. \ref{fig:exam2}, the Vitis analyzer compares the timing diagram of $example$ $2$ in both $Vitis$ and $Vitis$+$FastFlow$. The use of parallel threads in our $Vitis$+$FastFlow$ setup  accelerate the process flows.
\begin{table*}
\centering
\caption{Vitis vs FastFlow+Vitis}
\label{tab:comp}
\resizebox{16cm}{!}{
\setlength{\tabcolsep}{1mm} 
\def\arraystretch{1} 
\begin{tabular}{|M{0.4cm}|M{8cm}|p{5cm}|M{4cm}|M{3cm}|M{3cm}|P{3.25cm}|M{5cm}|M{2cm}|M{2cm}|M{2cm}|M{2cm}|}
  \hline
\large{\#}&  \large{\textbf{Examples} }  &   \shortstack[c]{\large{\textbf{Description}} \\ (hardware kernels are workers)}    &   \multicolumn{2}{c|}{\large{\textbf{Vitis}}} &   \multicolumn{3}{c|}{\textbf{\large{FastFlow+Vitis}}}& \multirow{2}{2cm}{\textbf{ \large{reduction of line \# in manual coding}}} & \multirow{3}{2cm}{ \large{\textbf{reduction of line \#in host.cpp}} }&\multirow{3}{2.5cm}{\large{\textbf{host.cpp generation time ($\mu s$)}}} &\multirow{3}{2cm}{\large{\textbf{kernels execution time($\mu s$)}}}\\\cline{4-8}

&      &      &   \large{\# lines in $host.cpp$ (\textbf{manual})} &  \large{\# lines \& definitions of static header file} & \large{\# lines in input dynamic codes (\textbf{manual})} & \large{ \# lines in $host.cpp$ (\textbf{automatic})}&\large{\# lines \&  definitions of static header files} &  & &&
  \\ \hline  
\large{1}&
\begin{tikzpicture}[node distance=0.6cm, scale=1.5, every node/.style={transform shape}, baseline=0]
\node(a)[rectangle, draw=black] {$E$};
\node(d)[rectangle, draw=gray, right of=a, right of =a, right of=a, above of=a] {$vadd\_1$};
\node(c)[rectangle, draw=gray, below of=d] {$vadd\_2$};
\node(e)[rectangle, draw=gray, right of=a, right of=a, right of=a, below of=a] {$vadd\_3$};
\node(f)[rectangle, draw=gray, below of=e] {$vadd\_4$};
\node(b)[rectangle, draw=black, right of=d, right of=d,right of=d, below of=d] {$C$};
\draw[->] (a) -- (d);
\draw[->] (a) -- (c);
\draw[->] (a) -- (e);
\draw[->] (a) -- (f);
\draw[->] (d) -- (b);
\draw[->] (c) -- (b);
\draw[->] (e) -- (b);
\draw[->] (f) -- (b);
\addvmargin{1mm}
\end{tikzpicture}   
&    \large{farm with 4 workers}   &\large{ $host.cpp=$\textbf{165} $connectivity=$\textbf{8}}&  & \large{$proc.csv$=\textbf{4} $circuit.csv$=\textbf{2}} &\large{\textbf{54}}&

\multirow{6}{5cm}{\large{$fdevice.hpp$ has \textbf{154} lines to define xclbin and kernel loading functions; $fnodetask.hpp$ has \textbf{59} lines to add dynamic pipeline stages for emitter, collector, middle \& FPGA node;
 $ftask.hpp$ has \textbf{183} lines to sync. inputs \& outputs of hardware kernels through memory;
  $nodes.cpp$ has \textbf{190} lines to defines logics of emitter, collector and middle node;
  $param.hpp$ has \textbf{38} lines to allocate base addr. of each CU;
  $hls\_stream.hpp$ has \textbf{30} lines for stream generation of inputs \& outputs of hardware kernels } }

& \large{\textbf{96 \%}}& \large{\textbf{67 \%}}&\large{$\sim$ 520}&\large{$\sim$ 1055}
  \\ \cline{1-4}  \cline{6-7} \cline{9-12}
\large{2}&\begin{tikzpicture}[node distance=0.4cm, scale=1.5, every node/.style={transform shape}, baseline=0]
\node(a)[rectangle, draw=black] {$E$};
\node(b)[rectangle, draw=gray, right of=a, right of =a, right of=a] {$vadd\_1$};
\node(c)[rectangle, draw=gray, right of=b, right of =b, right of=b] {$vmul\_1$};
\node(d)[rectangle, draw=gray, right of=c, right of =c, right of=c] {$vinc\_1$};
\node(e)[rectangle, draw=black, right of=d, right of =d, right of=d] {$C$};
\draw[->] (a) -- (b);
\draw[->] (b) -- (c);
\draw[->] (c) -- (d);
\draw[->] (d) -- (e);
\addvmargin{1mm}
\end{tikzpicture}    
  &    \large{3 pipes}   &   \large{$host.cpp=$\textbf{273} $connectivity=$\textbf{6}}& &  \large{$proc.csv$=\textbf{1}, $circuit.csv$=\textbf{6}}& \large{\textbf{36}}&& \large{\textbf{97 \%}}& \large{\textbf{86 \%}}& \large{$\sim$345}&\large{$\sim$ 301}
  \\ \cline{1-4}  \cline{6-7} \cline{9-12}
\large{3}& 
\begin{tikzpicture}[node distance=0.7cm, scale=1.3, every node/.style={transform shape}, baseline=0pt]
\node(a)[rectangle, draw=black] {$E$};
\node(d)[rectangle, draw=gray, right of=a, right of =a, above of=a] {$vadd\_1$};
\node(i)[rectangle, draw=gray, right of=d, right of=d] {$vmul\_1$};
\node(j)[rectangle, draw=gray, right of=i, right of=i] {$vinc\_1$};

\node(c)[rectangle, draw=gray, right of=a, right of=a] {$vadd\_2$};
\node(g)[rectangle, draw=gray, right of=c, right of=c] {$vmul\_2$};
\node(h)[rectangle, draw=gray, right of=g, right of=g] {$vinc\_2$};

\node(e)[rectangle, draw=gray, right of=a, right of=a,  below of=a] {$vadd\_3$};
\node(k)[rectangle, draw=gray, right of=e, right of=e] {$vmul\_3$};
\node(l)[rectangle, draw=gray, right of=k, right of=k] {$vinc\_3$};

\node(f)[rectangle, draw=gray, below of=e] {$vadd\_4$};
\node(m)[rectangle, draw=gray, right of=f, right of=f] {$vmul\_4$};
\node(n)[rectangle, draw=gray, right of=m, right of=m] {$vinc\_4$};

\node(b)[rectangle, draw=black, right of=d, right of=d,right of=d,  right of=d, right of=d, right of=d, below of=d] {$C$};
\draw[->] (a) -- (d);
\draw[->] (d) -- (i);
\draw[->] (i) -- (j);
\draw[->] (a) -- (c);
\draw[->] (c) -- (g);
\draw[->] (g) -- (h);
\draw[->] (a) -- (e);
\draw[->] (a) -- (f);
\draw[->] (j) -- (b);
\draw[->] (h) -- (b);
\draw[->] (l) -- (b);
\draw[->] (n) -- (b);
\draw[->] (e) -- (k);
\draw[->] (k) -- (l);
\draw[->] (f) -- (m);
\draw[->] (m) -- (n);
\addvmargin{1mm}
\end{tikzpicture}   
&\large{farm with 4 workers, each workers has 3 pipes} & \large{$host.cpp=$\textbf{286} $connectivity=$\textbf{24}} & \multirow{6}{2.6cm}{\large{$hls\_stream.hpp$ has \textbf{30} lines for stream generation of inputs \& outputs of hardware kernels}} & \large{$proc.csv$=\textbf{4}, $circuit.csv$=\textbf{6}}&\large{\textbf{80}}& & \large{\textbf{96 \%}}& \large{\textbf{72 \%}}&\large{$\sim$ 635}&\large{$\sim$ 3315}
  \\ \cline{1-4}  \cline{6-7} \cline{9-12}
\large{4}&\begin{tikzpicture}[node distance=0.6cm, scale=1.3, every node/.style={transform shape}, baseline=0pt]
\node(a)[rectangle, draw=black] {$E$};
\node(d)[rectangle, draw=gray, right of=a, right of =a, right of=a, above of=a] {$vadd\_1$};
\node(c)[rectangle, draw=gray, right of=d, right of =d, right of=d] {$vinc\_1$};
\node(e)[rectangle, draw=gray, right of=a, right of=a, right of=a, right of=a, right of=a] {$vmul\_1$};
\node(b)[rectangle, draw=black, right of=c, right of=c, right of=c, below of=c] {$C$};
\draw[->] (a) -- (d);
\draw[->] (d) -- (c);
\draw[->] (c) -- (b);
\draw[->] (a) -- (e);
\draw[->] (e) -- (b);
\addvmargin{1mm}
\end{tikzpicture}   &    \large{farm with 2 workers, 1$^{st}$ workers has 2 pipes \& 2$^{nd}$ worker has 1 pipe }& \large{$host.cpp=$\textbf{274} $connectivity=$\textbf{6}}&  & \large{$proc.csv$=\textbf{2}, $circuit.csv$=\textbf{6}}&  \large{\textbf{54}}&& \large{\textbf{97 \%}}& \large{\textbf{80 \%}}&\large{$\sim$ 494}&\large{$\sim$ 456}

  \\ \cline{1-4}  \cline{6-7} \cline{9-12}
\large{5}& 
\begin{tikzpicture}[node distance=0.65cm, scale=1.1, every node/.style={transform shape}, baseline=0pt]
\node(a)[rectangle, draw=black] {$E$};
\node(d)[rectangle, draw=gray, right of=a, right of =a, above of=a] {$vadd\_1$};
\node(i)[rectangle, draw=gray, right of=d, right of=d] {$vmul\_1$};

\node(c)[rectangle, draw=gray, right of=a, right of=a] {$vadd\_2$};
\node(g)[rectangle, draw=gray, right of=c, right of=c] {$vmul\_2$};

\node(e)[rectangle, draw=gray, right of=a, right of=a,  below of=a] {$vadd\_3$};
\node(k)[rectangle, draw=gray, right of=e, right of=e] {$vmul\_3$};

\node(b)[rectangle, draw=gray, right of=d, right of=d,right of=d,  right of=d,  below of=d] {$vinc_1$};
\node(f)[rectangle, draw=gray, right of=b, right of=b ] {$vinc_2$};
\node(h)[rectangle, draw=black, right of=f, right of=f ] {$C$};
\draw[->] (a) -- (d);
\draw[->] (d) -- (i);
\draw[->] (a) -- (c);
\draw[->] (c) -- (g);
\draw[->] (a) -- (e);
\draw[->] (i) -- (b);
\draw[->] (k) -- (b);
\draw[->] (e) -- (k);
\draw[->] (g) -- (b);
\draw[->] (b) -- (f);
\draw[->] (f) -- (h);
\addvmargin{1mm}
\end{tikzpicture}   
&\large{farm with 3 workers, each workers has 2 pipes. these two workers connected with 2 common pipes} &  \large{$host.cpp=$\textbf{276} $connectivity=$\textbf{16}}& & \large{$proc.csv$=\textbf{4}, $circuit.csv$=\textbf{6}}&\large{\textbf{80}}& & \large{\textbf{96 \%}}& \large{\textbf{71 \%}}&\large{$\sim$ 230}&\large{$\sim$ 714} \\ \hline
 \multicolumn{12}{|c|}{}\\
 \multicolumn{12}{|c|}{\textbf{\large{All five designs developed by $FastFlow$+$Vitis$ and $Vitis$ consume the same power, critical path, and slices.}}}
  \\ \hline
  \end{tabular}
  }
\end{table*}

\vspace{-6pt}
\section{Conclusion}
\label{sec:con}
Integrating $FastFlow$ with $Vitis$ drastically reduces the code-writing effort for parallel programming across FPGA Stacks in data centers while still achieving the same design performance as $Vitis$.
This work describes the preliminary results of our hybrid tool flow with $FastFlow$ and $Vitis$ libraries. The proposed script can automate the placement of hardware tasks in multiple FPGAs of data centers with appropriate data synchronization of inputs and outputs of hardware tasks. This proposed framework is scalable for a stacks of FPGAs, and it automatically handles data synchronization of parallel hardware kernels running in different FPGAs. In the future, this proposed tool flow will try to add other complex farm and pipe structures. 
Including input and output slots in $proc.csv$ of our future version of tool flow can offer more customization in farm and pipe structures. The code and test cases of this work is uploaded to GitHub \footnote{\label{git} https://github.com/rourabpaul1986/fastflow\_fpga\_stacks}. This research is supported by multiple funding agencies \footnote{\label{ack} This work is partially supported by Spoke 1 "Future HPC \& Big Data" of the Italian Research Center on High-Performance Computing, Big Data and Quantum Computing (ICSC) funded by MUR Missione 4 Componente 2 Investimento 1.4: Potenziamento strutture di ricerca e creazione di "campioni nazionali di R\&S (M4C2-19 )" - Next Generation EU (NGEU).}.
\bibliographystyle{unsrt}  
\bibliography{IEEEexample}

\end{document}